\documentclass[conference,a4paper]{IEEEtran}
\IEEEoverridecommandlockouts
\usepackage{cite}
\usepackage{amsmath,amssymb,amsfonts}
\usepackage{algorithmic}
\usepackage{graphicx}
\usepackage{textcomp}
\usepackage{xcolor}
\usepackage[nolist]{acronym}
\usepackage{paralist}
\usepackage{babel}
\usepackage{siunitx}
\usepackage[absolute]{textpos}

\begin{document}

\makeatletter
\def\ps@IEEEtitlepagestyle{%
\def\@oddfoot{\parbox{\textwidth}{\footnotesize
Author's version of a paper accepted for publication in Proceedings of 2021 IEEE PES Innovative Smart Grid Technologies Europe (ISGT-Europe).\\[1em]
\textcopyright{} 2021 IEEE. 
Personal use of this material is permitted.  
Permission from IEEE must be obtained for all other uses, in any current or future media, including reprinting/republishing this material for advertising or promotional purposes, creating new collective works, for resale or redistribution to servers or lists, or reuse of any copyrighted component of this work in other works.\vspace{1.2em}}
}%
}
\makeatother

\begin{acronym}
\acro{sg}[SG]{smart grid}
\acroplural{sg}[SGs]{smart grids}
\acro{der}[DER]{distributed energy resource}
\acroplural{der}[DERs]{distributed energy resources}
\acro{ict}[ICT]{information and communication technology}
\acro{fdi}[FDI]{false data injection}
\acro{scada}[SCADA]{Supervisory Control and Data Acquisition}
\acro{mtu}[MTU]{Master Terminal Unit}
\acroplural{mtu}[MTUs]{Master Terminal Units}
\acro{hmi}[HMI]{Human Machine Interface}
\acro{plc}[PLC]{Programmable Logic Controller}
\acroplural{plc}[PLCs]{Programmable Logic Controllers}
\acro{ied}[IED]{Intelligent Electronic Device}
\acroplural{ied}[IEDs]{Intelligent Electronic Devices}
\acro{rtu}[RTU]{Remote Terminal Unit}
\acroplural{rtu}[RTUs]{Remote Terminal Units}
\acro{iec104}[IEC-104]{IEC 60870-5-104}
\acro{apdu}[APDU]{Application Protocol Data Unit}
\acro{apci}[APCI]{Application Protocol Control Information}
\acro{asdu}[ASDU]{Application Service Data Unit}
\acro{io}[IO]{information object}
\acroplural{io}[IOs]{information objects}
\acro{cot}[COT]{cause of transmission}
\acro{mitm}[MITM]{Man/Machine-in-the-Middle}
\acro{fdi}[FDI]{False Data Injection}
\acro{ids}[IDS]{intrusion detection systems}
\acro{mv}[MV]{medium voltage}
\acro{lv}[LV]{low voltage}
\acro{cdss}[CDSS]{controllable distribution secondary substation}
\acro{bss}[BSS]{battery storage system}
\acroplural{bss}[BSSs]{battery storage systems}
\acro{pv}[PV]{photovoltaic inverter}
\acro{mp}[MP]{measuring point}
\acroplural{mp}[MPs]{measuring points}
\acro{dsc}[DSC]{Dummy SCADA Client}
\acro{fcli}[PVI2]{PV Inverter 2}
\acro{fipi}[PVI1]{PV Inverter 1}
\acro{sii}[BSSI]{Battery Storage System Inverter}
\acro{tls}[TLS]{Transport Layer Security}
\acro{actcon}[ActCon]{Activation Confirmation}
\acro{actterm}[ActTerm]{Activation Termination}
\acro{rtt}[RTT]{Round Trip Time}
\acro{c2}[C2]{Command and Control}
\end{acronym}

\title{Investigating Man-in-the-Middle-based False Data Injection in a Smart Grid Laboratory Environment}

\author{
\IEEEauthorblockN{%
Ömer Sen\IEEEauthorrefmark{1},
Dennis van der Velde\IEEEauthorrefmark{1},
Philipp Linnartz\IEEEauthorrefmark{2},
Immanuel Hacker\IEEEauthorrefmark{1}
}
\IEEEauthorblockN{%
Martin Henze\IEEEauthorrefmark{3},
Michael Andres\IEEEauthorrefmark{1},
Andreas Ulbig\IEEEauthorrefmark{2}
}

\IEEEauthorblockA{%
\IEEEauthorrefmark{1}\textit{Digital Energy, Fraunhofer FIT,} Aachen, Germany\\
Email: \{oemer.sen, dennis.van.der.velde, immanuel.hacker, michael.andres\}@fit.fraunhofer.de}

\IEEEauthorblockA{%
\IEEEauthorrefmark{2}\textit{High Voltage Equipment and Grids, Digitalization and Energy Economics, RWTH Aachen University,} Aachen, Germany\\
Email:  \{philipp.linnartz, andreas.ulbig\}@iaew.rwth-aachen.de}

\IEEEauthorblockA{%
\IEEEauthorrefmark{3}\textit{Cyber Analysis \& Defense, Fraunhofer FKIE,} Wachtberg, Germany\\
Email: martin.henze@fkie.fraunhofer.de}
}

\IEEEoverridecommandlockouts
\IEEEpubid{\makebox[\columnwidth]{~\hfill} \hspace{\columnsep}\makebox[\columnwidth]{ }}

\maketitle

\IEEEpubidadjcol

\begin{abstract}
With the increasing use of information and communication technology in electrical power grids, the security of energy supply is increasingly threatened by cyber-attacks.
Traditional cyber-security measures, such as firewalls or intrusion detection/prevention systems, can be used as mitigation and prevention measures, but their effective use requires a deep understanding of the potential threat landscape and complex attack processes in energy information systems.
Given the complexity and lack of detailed knowledge of coordinated, timed attacks in smart grid applications, we need information and insight into realistic attack scenarios in an appropriate and practical setting. %
In this paper, we present a man-in-the-middle-based attack scenario that intercepts process communication between control systems and field devices, employs false data injection techniques, and performs data corruption such as sending false commands to field devices.
We demonstrate the applicability of the presented attack scenario in a physical smart grid laboratory environment and analyze the generated data under normal and attack conditions to extract domain-specific knowledge for detection mechanisms.
\end{abstract}


\section{Introduction} \label{sec:introduction}
The ongoing transformation of electric power systems in \acp{sg} is driven primarily by the increasing penetration of volatile \acp{der} and new load situations created by prosumer entities~\cite{1_tuballa2016review}.
Thus, grid operators face new challenges, including congestion and bidirectional power flow situations, which require timely active grid operation through the increased use of \ac{ict}~\cite{3_van2020methods,klaer_sgam_2020}.
Consequently, the increasing dependence on the \ac{ict} infrastructure and system complexity therefore generates new threats.
For example, the secure operation of the power grid is at risk due to incorrect configuration of communication links, faulty grid automation algorithms, or cyber attacks~\cite{3_van2020methods,6_case2016analysis}.
In this new threat landscape for \acp{sg}, cyber-attack vectors that exploit the vulnerabilities of advanced \ac{ict} infrastructures occupy a central position in power supply threat scenarios~\cite{1_tuballa2016review,8_kotut2016survey,9_mendel2017smart}.
Severe cyber-attacks aim to compromise the integrity, confidentiality, and availability of grid operations to distort operational decisions, obtain sensitive information, or delay or disrupt the functioning of services~\cite{10_li2012securing}.
As cyber-attacks become more diverse and sophisticated~\cite{9_mendel2017smart}, and information and operational technologies continue to converge, novel cross-domain security strategies are needed to ensure a reliable power supply.
For instance, \ac{ids}, next-generation firewalls, role-based access control, and encryption methods aid to address system vulnerabilities, detect various cyber-attacks, take the appropriate countermeasures, and identify the entities involved within the attack~\cite{henze_blueprint_2020}.
The development and validation of such cross-domain security strategies depends on the availability and quality of information from all relevant attack scenarios~\cite{17_zuech2015intrusion}.
However, insights and information from real cyber incidents in power grids that provide characteristics and signatures of the attack are limited to the scientific community for privacy and security reasons~\cite{8_kotut2016survey} and are usually generated by studying synthetically replicated attack scenarios~\cite{17_zuech2015intrusion}. %
Since it is difficult to replicate attack scenarios that have severe consequences in a productive, operating power grid for investigation purposes, an isolated, secure and controllable environment is required that provides valid properties not only within the energy domain but also of the \ac{ict}.
Consequently, a test environment of an energy information system with its primary and secondary technology components is required, in which attack scenarios affecting both the \ac{ict} and the electrical grid are deployed and investigated to extract useful characteristics for the development of appropriate countermeasures. %
Therefore, in order to gain useful insights from experiments that can be used to develop and validate data-driven countermeasures such as \ac{ids}, detailed documentation of the underlying infrastructure and attack scenarios, including both \ac{ict} and process views, is essential for meaningful adoption of experiment results for security measures.
In this paper, we present a structured and comprehensive approach to conduct experiments under normal operating and attack conditions in a cyber-physical \ac{sg} lab environment and gain useful knowledge for countermeasures such as process-aware \ac{ids} from a targeted analysis.
The structure as well as the contribution of this paper are:
\begin{enumerate}
    \item We provide an overview of relevant cyber-security issues in \acp{sg}, as well as current challenges in legacy-compliant security measures for traditional power grids (Section~\ref{sec:background}).
    \item We show and describe a laboratory environment that replicates an \ac{sg} use case with high \ac{ict} penetration for conducting cyber-security investigations (Section~\ref{sec:testbed}). %
    \item We present and discuss a structured and traceable setup for performing \ac{mitm}-based \ac{fdi} attacks in \ac{sg} lab environments to gain insights into complex attack techniques (Section~\ref{sec:attack}).
    \item We demonstrate the feasibility of the developed attack tool and investigate communication- and process-level characteristics for detecting such attacks using novel \ac{ids} approaches. (Section~\ref{sec:result}).
\end{enumerate}

\section{Cyber Security in Smart Grids} \label{sec:background}
As a basis for our work, we provide a brief overview of the core components and their security issues in \ac{sg} utilities and highlight existing challenges in prominent security measures.

\subsection{Cyber Security in Power Process Networks} \label{subsec:background_cybersec}
Process networks in \acp{sg} provide interconnectivity between control centers, field devices, and assets on which \ac{scada} systems are responsible for monitoring and controlling automatic operations in transmission or a distribution substation~\cite{18_radoglou2019attacking}.
In particular, communication standardized via industrial protocols such as \ac{iec104} or DNP3 is carried out via \acp{mtu} with the logical controllers such as \acp{rtu} and \acp{ied}, which in turn are responsible for monitoring the processes in the industrial environment by interacting with the sensors and actuators.
Initially designed for use in narrowly defined networks, \ac{iec104} as a legacy industrial protocol does not provide any security functionalities, in particular neither encryption nor authentication~\cite{20_wg15iec}.
This means that process data is transmitted unencrypted, allowing unauthorized third parties to monitor and analyze data traffic (e.g., reading out certain memory locations), which can have severe consequences for the provision of a reliable power supply.
In addition, many commands of the protocol, such as reset command, query commands, and read commands, do not have authentication mechanisms built in, allowing unauthorized access.
One exploit of the aforementioned protocol vulnerability would be the \ac{mitm}-based \ac{fdi}, which refers to an attacker who could intercept the logical connection between communicating devices (e.g., between \ac{mtu} and \ac{rtu})~\cite{serror_iiot_2020} and inject their messages (e.g., a false command) through an \ac{mitm} attack~\cite{22_rodofile2018generating}.
Given the lack of security mechanisms in the \ac{iec104} protocol, the attacker would then be able to read, modify, inject, or discard sent or new messages between the intercepted endpoints~\cite{23_yang2012man}.

\subsection{Security Measures and Challenges} \label{subsec:background_securitymeasures}
In view of the prominent security issues in process networks, various research and discussions are being conducted to secure \ac{iec104} communications based on the security principles of the IEC 62351 standard~\cite{21_todeschini2020securing}.
In particular, the integration of cipher suites, as used in the \ac{tls} protocol, is mandated by the IEC 62351 standard to secure end-to-end communication between two connection terminals through secure key exchange, encryption, and authentication~\cite{31_iec62351_2018}.
However, traditional process networks usually consist of performance-limited devices, so the trade-off between security and performance can result in critical real-time requirements not being met~\cite{32_tanveer2020secure}.
An approach which does not actively interfere with the process network is an \ac{ids}, which typically monitors network traffic with passive sniffing and detects attack indicators over the monitored network~\cite{37_fernandes2019comprehensive}.
Known detection methods in this area include identifying attack indicators by comparing captured traces to a specification or model representing either system behavior under normal operating conditions (i.e., anomaly-based) or attack scenarios (i.e., misuse-based), with any discrepancy or match resulting from this comparison leading to the generation of an alert~\cite{17_zuech2015intrusion}.
However, these detection approaches require a comprehensive, robust, and diverse data foundation and understanding in terms of attack signatures from real-world cyber incidents or system behavior under normal operating conditions~\cite{38_khraisat2019survey}.
In this paper, we present a \ac{mitm}-based \ac{fdi} attack scenario in a real laboratory setup and study the observations during the experiment to provide useful insights and information that can be used within a process-aware detection mechanism to effectively detect such attacks.

\subsection{Related Work} \label{subsec:background_relatedwork}
The security issues regarding the study of the impact of \ac{mitm}-based \ac{fdi} attacks in \acp{sg} have been investigated in several directions. %
In particular, research has been conducted in this regard to investigate \ac{fdi}-based attack scenarios in narrowly defined power system testbeds~\cite{24_adepu2018epic}, also emulating \ac{mitm}-based attacks for \ac{iec104} security and risk assessments~\cite{18_radoglou2019attacking}. %
Moreover, further work is being done to present an approach to providing a data generation framework focused on DNP3 \ac{mitm}-based injection attacks~\cite{25_rodofile2017framework} and data and control manipulation attacks over IEC 61850 in secondary substations~\cite{26_biswas2019synthesized}. %
In addition, studies have been conducted on synthesis frameworks for specific attack vectors~\cite{27_babu2017melody} as well as on an automated marking process for deployed protocol-specific attack~\cite{22_rodofile2018generating}.
In this context, research on stealthy \ac{mitm} attacks for \ac{fdi} in DNP3 or Profinet is also conducted in a cyber-physical test environment to analyze the impact on latency~\cite{39_wlazlo2021man} or generate datasets for data-driven detection approaches~\cite{40_noorizadeh2021cyber}.
Many of the related works involve the study of cyber-attacks on power grids for data generation and consequence analysis.
However, most of them lack sufficient coverage of the generated data within the experiments themselves, especially from a process perspective, and the majority of experiments are conducted in a (partially) virtualized testbed. %
In this work, we provide insight and analysis of \ac{mitm}-based \ac{fdi} attack experiments in terms of their \ac{ict} and process data components performed in a physically deployed \ac{iec104} \ac{sg} laboratory environment consisting of typically used components. %

\section{Laboratory Environment} \label{sec:testbed}
To replicate \ac{mitm}-based \ac{fdi} attack scenarios, we present and describe an \ac{sg} lab environment and experiment setup in this section. %
Due to the homogeneous structure of process networks in the energy sector and the selectively chosen assets in our testbed setup, our experimental results provide high applicability for security measures in this domain.
Our experiment aims to cover later stages of cyber-attacks, assuming that physical access to the network perimeter has already been gained in worst-case scenarios.

\subsection{Smart Grid Testbed Setup} \label{subsec:testbed_testbed}
\begin{figure}
    \centerline{\includegraphics[width=\columnwidth]{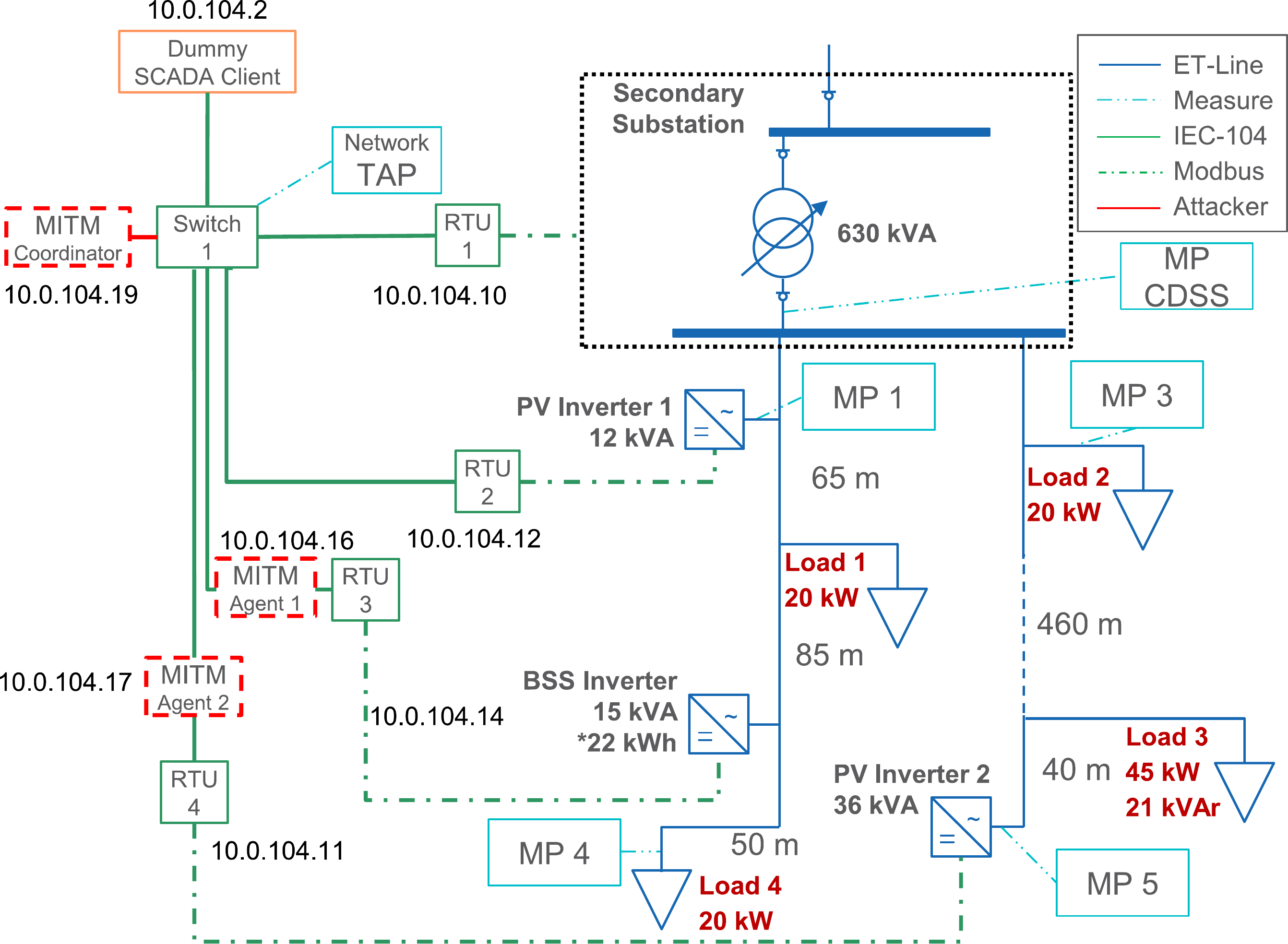}}
    \caption{Our \ac{sg} testbed setup replicates an \ac{mv}/\ac{lv} distribution grid and consists of multiple \ac{der} assets, resistive/inductive loads, and a controlling field device within a \ac{scada} network.}
    \label{fig:tesbed_setup}
    \vspace{-1em}
\end{figure}

The \ac{sg} testbed replicates an \ac{mv} / \ac{lv} distribution grid equipped with networked assets.
All components in our cyber-physical testbed are physically deployed with no virtual simulation of equipment.
We present the underlying \ac{ict} and electrical topologies in Figure~\ref{fig:tesbed_setup}.
Within the testbed setup, the electrical assets consist of a 10 kV / 0.4 kV \ac{cdss} with a 630 kVA transformer, 22 kWh \ac{bss}, 12 kVA and 36 kVA \acp{pv}, and several resistive/inductive loads.
The \ac{lv} grid is configured with a radial topology consisting of two strings, the first of which is equipped with a 200 m, and resp. 500 m NAYY cable. %
In addition, the \ac{mv}/\ac{lv} grid is equipped with five integrated three-phase current and voltage \acp{mp} with built-in power analyzers for current and voltage measurements as well as power quality measurements (e.g., harmonics).
The \ac{der} assets are commercial inverters with adjustable feed-in profiles and programmable reactive power control (Q(U), Q(P), cos $\phi$) fed from DC sources.
These assets are controlled and monitored via \acp{rtu} with Modbus.
In addition, the testbed's process network consists of an \ac{ict} switch that connects the \acp{rtu} to the \ac{dsc}.
In this regard, the \ac{dsc} in our testbed is replicated by an \ac{iec104} \ac{mtu} that sends control and query commands to \acp{rtu}.
Consequently, our testbed setup provides realistic behavior of secondary and primary technology components in terms of communication and electrical energy exchange, enabling cross-domain investigation of \ac{mitm}-based \ac{fdi} attack scenarios.

\subsection{Experiment Setup} \label{subsec:testbed_expirement}
Within this work, the described testbed will be the subject of \ac{mitm}-based \ac{fdi} attack experiments performed with a specially developed \ac{mitm} tool (cf. Section~\ref{sec:attack}).
As shown in Figure~\ref{fig:tesbed_setup}, two \ac{mitm} agents are placed to intercept one or more communication channels, i.e., placed between \ac{rtu} 3 and Switch 1 and between Switch 1 and \ac{rtu} 4.
To capture the traffic of different communication channels within the process network, we also place network taps on the \ac{ict} switch and \ac{mitm} agents.
The \ac{mitm} coordinator is directly connected to Switch 1 and uses the existing \ac{ict} infrastructure of the testbed to communicate with the \ac{mitm} agents.
We perform the following experiment: %
\begin{enumerate}[(a)]
\item Testbed operation with \ac{mitm} deployment, but no active interference with systems.
During this phase of the experiment ($t<\num{500}\si{\second}$), the \ac{dsc} sends stepwise power reduction control commands with set-points of \num{30}\% and \num{50}\% nominal power to the \ac{fcli} via \ac{rtu} 4 at approximately \num{360}\si{\second}.
At about \num{480}\si{\second}, the \ac{dsc} sends 21\% and 42\% of the rated power discharge command to the \ac{sii} via \ac{rtu} 3.
\item Testbed operation with an active \ac{mitm} performing \ac{fdi}.
Within this experimentation phase ($t>\num{500}\si{\second}$), the \ac{mitm} coordinator sends action packets to \ac{mitm} Agent 2, instructing it to send a set-point command of \num{100}\% at about $\num{600}\si{\second}$ to the \ac{fcli}.
\ac{mitm} Agent 1 is instructed at approximately $\num{750}\si{\second}$ to issue a control command with a set-point of \num{-42}\% nominal power to the \ac{sii}.
\end{enumerate}

\section{Man-in-the-Middle-based False Data Injection} \label{sec:attack}
\begin{figure}[t]
    \centerline{\includegraphics[width=\columnwidth]{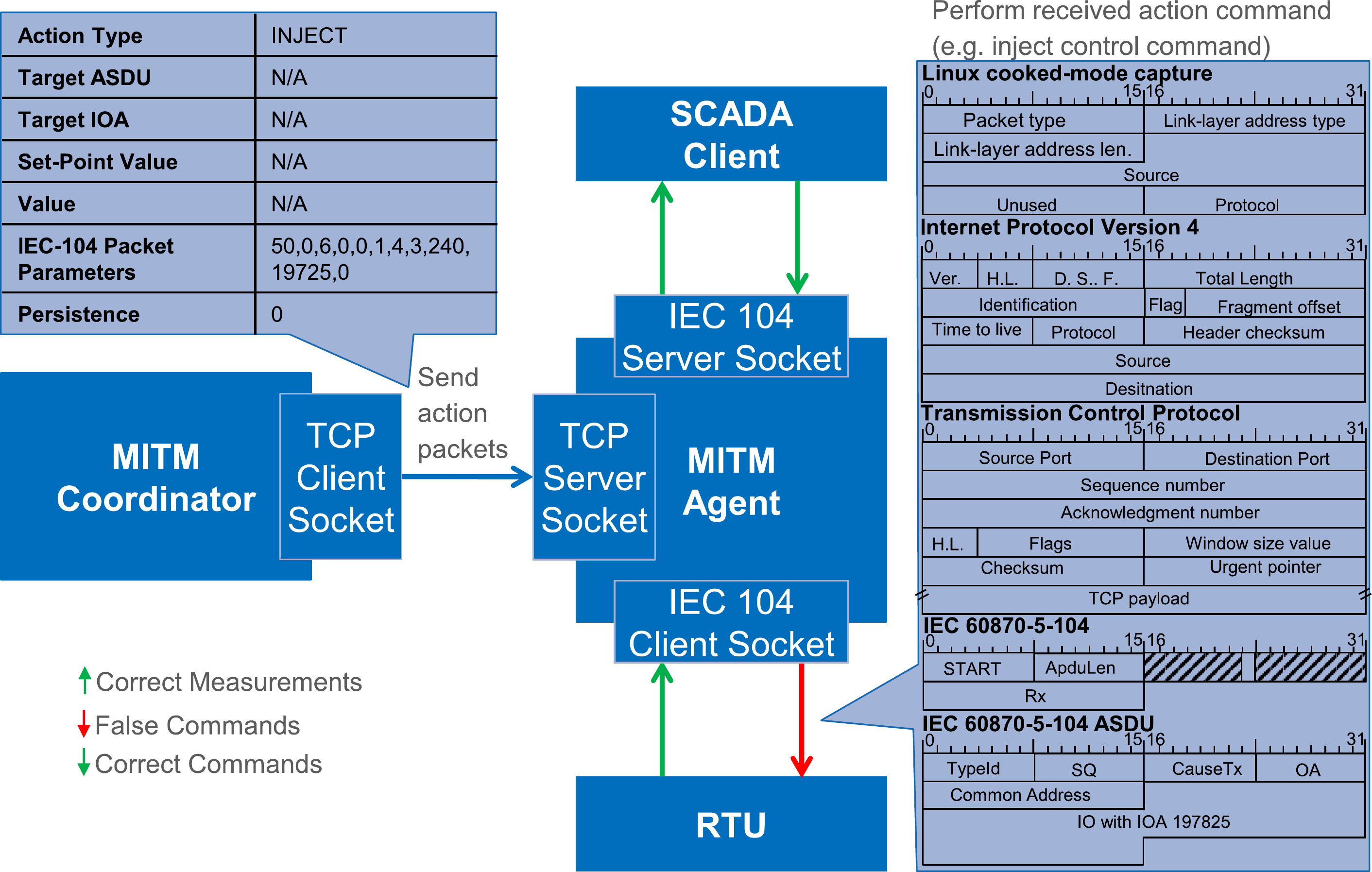}}
    \caption{The \ac{mitm} agent sits on the communication path between a \ac{rtu} and the \ac{mtu} (e.g., on a compromised network switch) and is instructed by the \ac{mitm} coordinator to perform actions such as manipulating, injecting, or collecting transmitted or new data.} %
    \label{fig:mitm_composition}
    \vspace{-1em}
\end{figure}
\begin{figure}[t]
    \centerline{\includegraphics[width=\columnwidth]{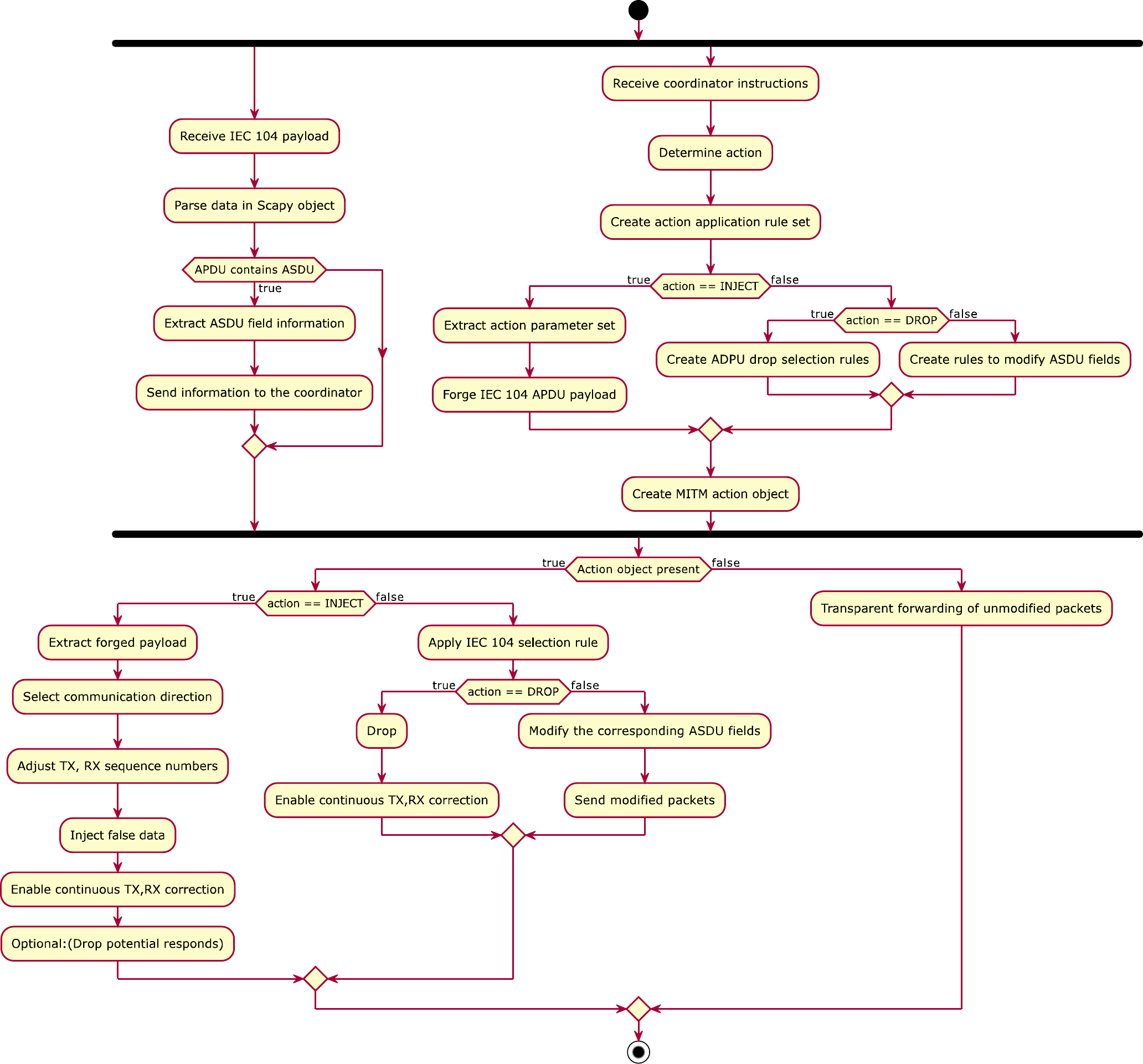}}
    \caption{High-level representation of \ac{mitm} agent logic interacting with \ac{mitm} coordinator to execute remotely received instructions.}
    \label{fig:mitm_logic}
    \vspace{-1em}
\end{figure}
As a basis for our \ac{fdi} attack experiments, we developed a special \ac{mitm} tool for \ac{iec104} for transparent physical interception.
The implementation is based on worst-case assumptions, where the attacker surreptitiously intercepts the communication on already layer 1.
Our \ac{mitm}-attack tool, as shown in Figure~\ref{fig:mitm_composition}, follows a coordinator-agent architecture, where the coordinator remotely controls the behavior of agents (located between two target endpoints) via certain action packets (cf. Figure~\ref{fig:mitm_logic}).
The action packets received via TCP connections are used to determine which actions the agent has to perform (e.g., manipulate measured values or inject new control commands).
They also contain application rules, such as for the inject action a collection of parameters needed for local \ac{iec104} package forgery.
To bridge the target communication channel, the agent must operate as an inline bridge~\cite{30_bohme2000linux}, i.e., take a hybrid role and provide counterpoints to the respective \ac{mtu} and \ac{rtu} endpoints to establish and maintain the connection and perform transparent proxying. %
The design of our \ac{mitm} tool is based on transparent proxying to enable consistent protocol-compliant and selective forwarding of intercepted traffic.
Therefore, our \ac{mitm} tool intercepts the connection between the \ac{mtu} and \ac{rtu} endpoints via a virtual network bridge based on a Linux operating system using the bridge utilities.
Since our \ac{mitm} agent intercepts directly on a Layer 1 basis, we assign the agent's two physical network interfaces, the incoming client-side interface and the outgoing origin-server-side interface, to the newly created bridge interface.
The bridged interfaces share an IP address provided by the bridge interface, which the \ac{mitm} coordinator uses to reach the \ac{mitm} agent.
Then, to selectively lift \ac{iec104} traffic for the \ac{fdi}, incoming packets on a bridged interface that are to be forwarded to another bridged interface are instead intercepted and forwarded through the BROUTING chain of ebtables when the \ac{iec104} protocol is associated. %
Thus, when \ac{iec104} traffic is detected in the incoming packets arriving at a (forwardable) interface, the packet flow is marked and policy routing is used depending on the directionality of the bridged interfaces.
In the following, transparent proxying via TPROXY for inbound packets is needed to force the acceptance of packets to foreign IP addresses, i.e., intercept connections between clients (\acp{mtu}) and servers (\acp{rtu}) without being visible.
Hereby, inbound transparency describes being transparent to connections that enter the proxy, while outbound transparency describes being transparent to connections that originate from the proxy.
To allow our \ac{fdi} application to access the \ac{iec104} packets, the marked traffic is routed back to the localhost application of the \ac{mitm} agent via a policy routing table through the loopback interface. %
In the Python-based \ac{fdi} application, transparent TCP sockets are used to read the forwarded \ac{iec104} data, even if this data is not intended for the host itself.
Consequently, to intercept the communication channel between the \ac{rtu} and \ac{mtu} endpoints, the \ac{mitm} agent opens the sockets as a counter endpoint for connection establishment by first providing a server-side socket that listens for connection requests from the \ac{mtu}.
When the connection between the \ac{mtu} and the \ac{mitm} agent is established, the \ac{mitm} agent acts as a client and continuously sends connection requests to the \ac{rtu}.
If the connection to the \ac{rtu} is successful, the \ac{mitm} agent forwards the data internally to the \ac{mtu} passing through the \ac{mitm} agent application.
In this phase, Scapy~\cite{29_biondi2005packet} is used to manipulate the intercepted \ac{iec104} communications, allowing selective high-level manipulation of the \ac{asdu} payload by parsing the packet bytes into an accessible structured Python object and converting them back into bytes for forwarding to the appropriate destination endpoints.
In addition, the injection of new \ac{iec104} packets is also based on Scapy, which allows modular, protocol-compliant packet forgery according to external parameter sets.
However, actively changing the payload size of packets or transmitted frames in the communication channel, e.g., for inject or drop actions, requires stateful continuous correction of certain protocol fields such as checksum or sequence numbers.
By converting the socket instances into Scapy socket streams, no adaptation within the underlying Ethernet/IP/TCP protocol layers is required, as Scapy takes care of the consistent correction.
However, at the \ac{iec104} application level, the transmitted - TX, received - RX sequence numbers within the \ac{apci} must be continuously corrected, depending on the security functions within the \ac{iec104} endpoint applications.
This correction is done at the level of the \ac{fdi} script, where the amount of sent and received \ac{apdu} packets of each communication direction is counted and adjusted according to the amount of injected or discarded packets.
Based on our approach, we are able to replicate \ac{c2}-based attack scenarios to deploy distributed \ac{mitm}-based agents that perform transparent \ac{iec104} communication interception with remote control capabilities, forming the basis for coordinated attacks.

\section{Analysis for the Detection of MITM-Based FDI} \label{sec:result}
\begin{figure}[t]
    \centerline{\includegraphics[width=\columnwidth]{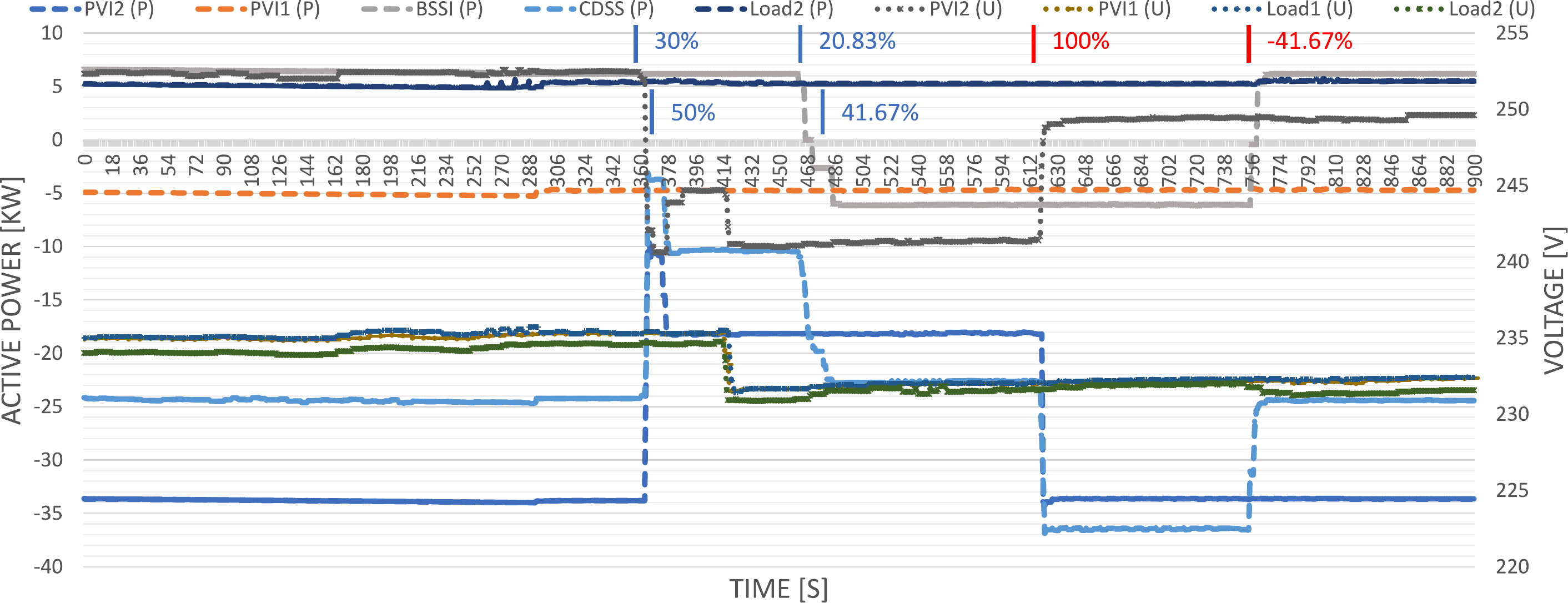}}
    \caption{Recorded power and voltage measurements at \ac{mp} sites and commands sent by \ac{dsc} and \ac{mitm} agents are shown, with the actions of \ac{fdi} occurring at \num{600}\si{\second} with \num{100}\% and \num{750}\si{\second} with \num{-41.67}\% set-point value.}
    \label{fig:result_plot_measurements}
    \vspace{-1em}
\end{figure}
\begin{figure}[t]
    \centerline{\includegraphics[width=\columnwidth]{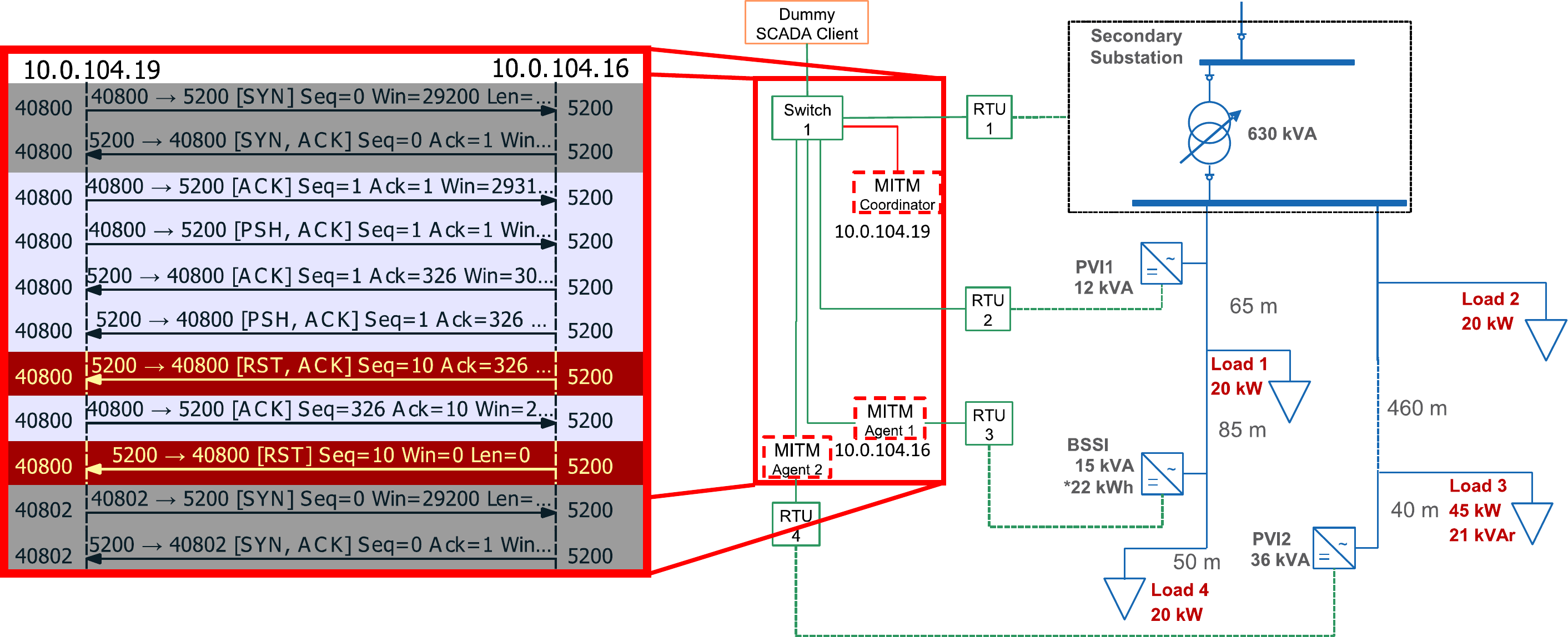}}
    \caption{Flowchart of observed communication within the \ac{c2} infrastructure between \ac{mitm} coordinator (10.0.104.19) and \ac{mitm} agent (10.0.104.16).}
    \label{fig:result_plot_c2}
    \vspace{-1em}
\end{figure}
\begin{figure}[t]
    \centerline{\includegraphics[width=\columnwidth]{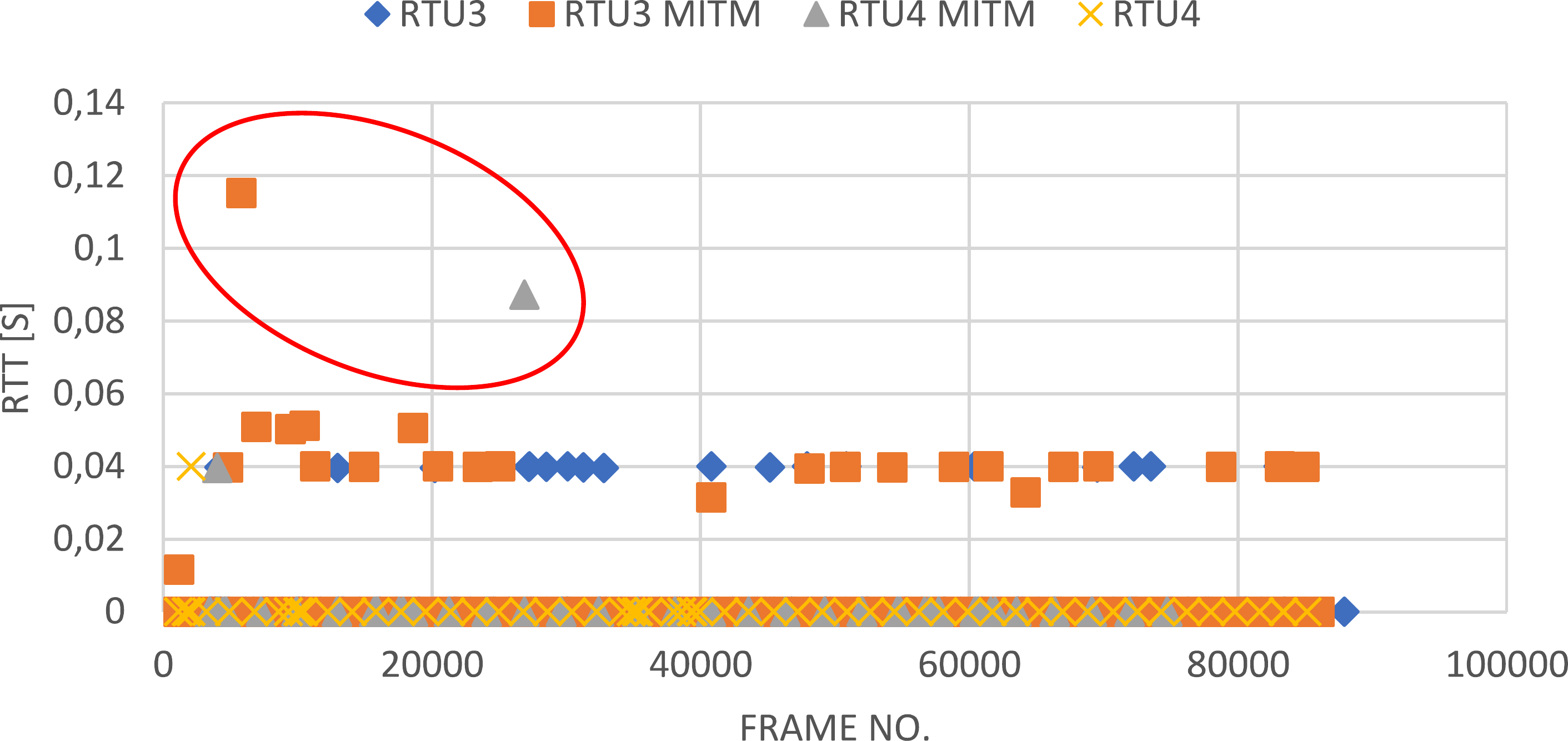}}
    \caption{\ac{rtt} of \ac{iec104} traffic originating from \acp{rtu} 3 and 4, with and without interception by \ac{mitm} agents, showing significant latency outliers.}
    \label{fig:result_plot_rtt}
    \vspace{-1em}
\end{figure}
\begin{figure}[t]
    \centerline{\includegraphics[width=\columnwidth]{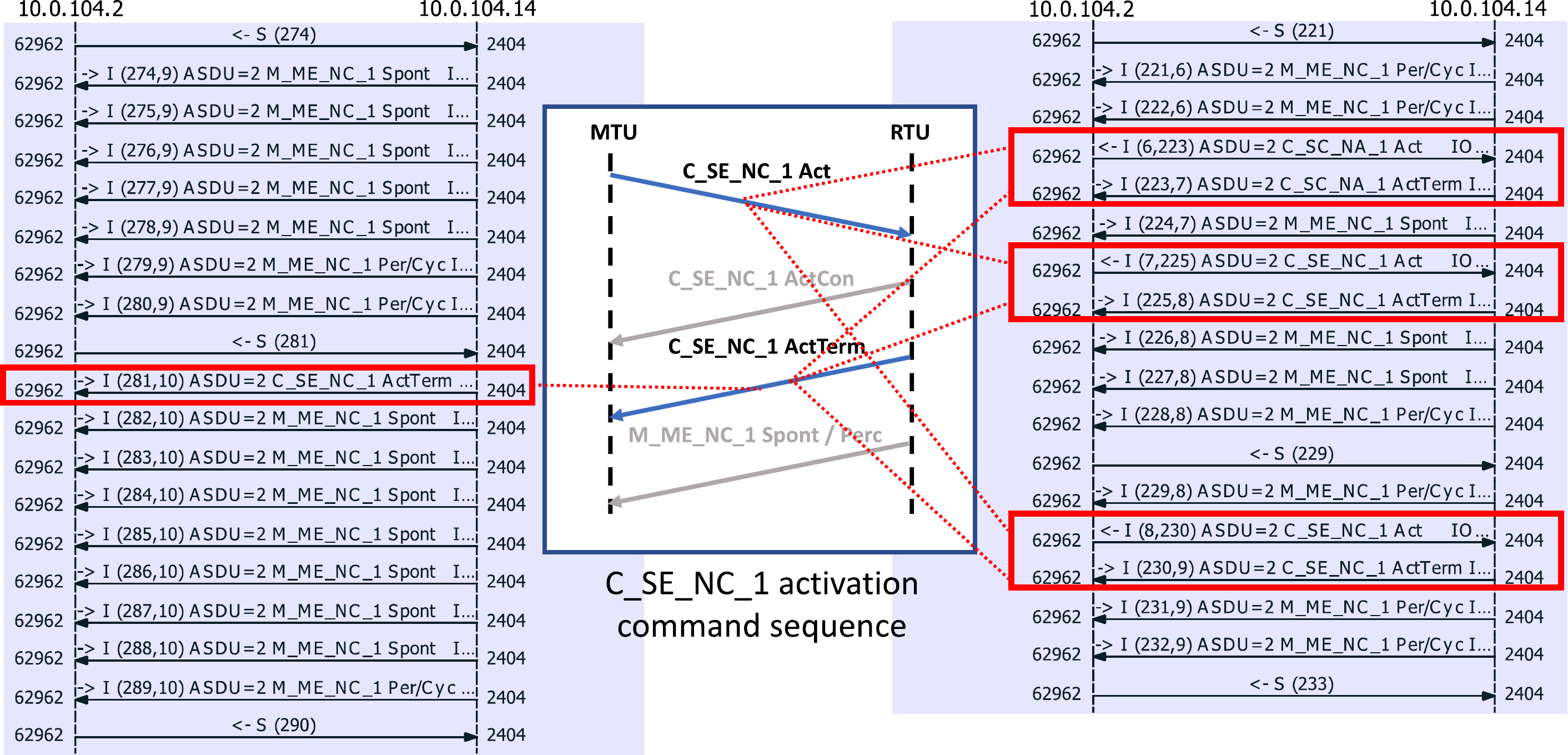}}
    \caption{Illustration of the sequence of activation of the set-point command C\_SE\_NC\_1 between \ac{dsc} and \ac{rtu} 3 with \ac{mitm}-based \ac{fdi} (left) and the normal operation of \ac{dsc} (right), showing anomalous flow behavior.}
    \label{fig:result_plot_cmdflow}
    \vspace{-1em}
\end{figure}
\begin{figure}[t]
    \centerline{\includegraphics[width=\columnwidth]{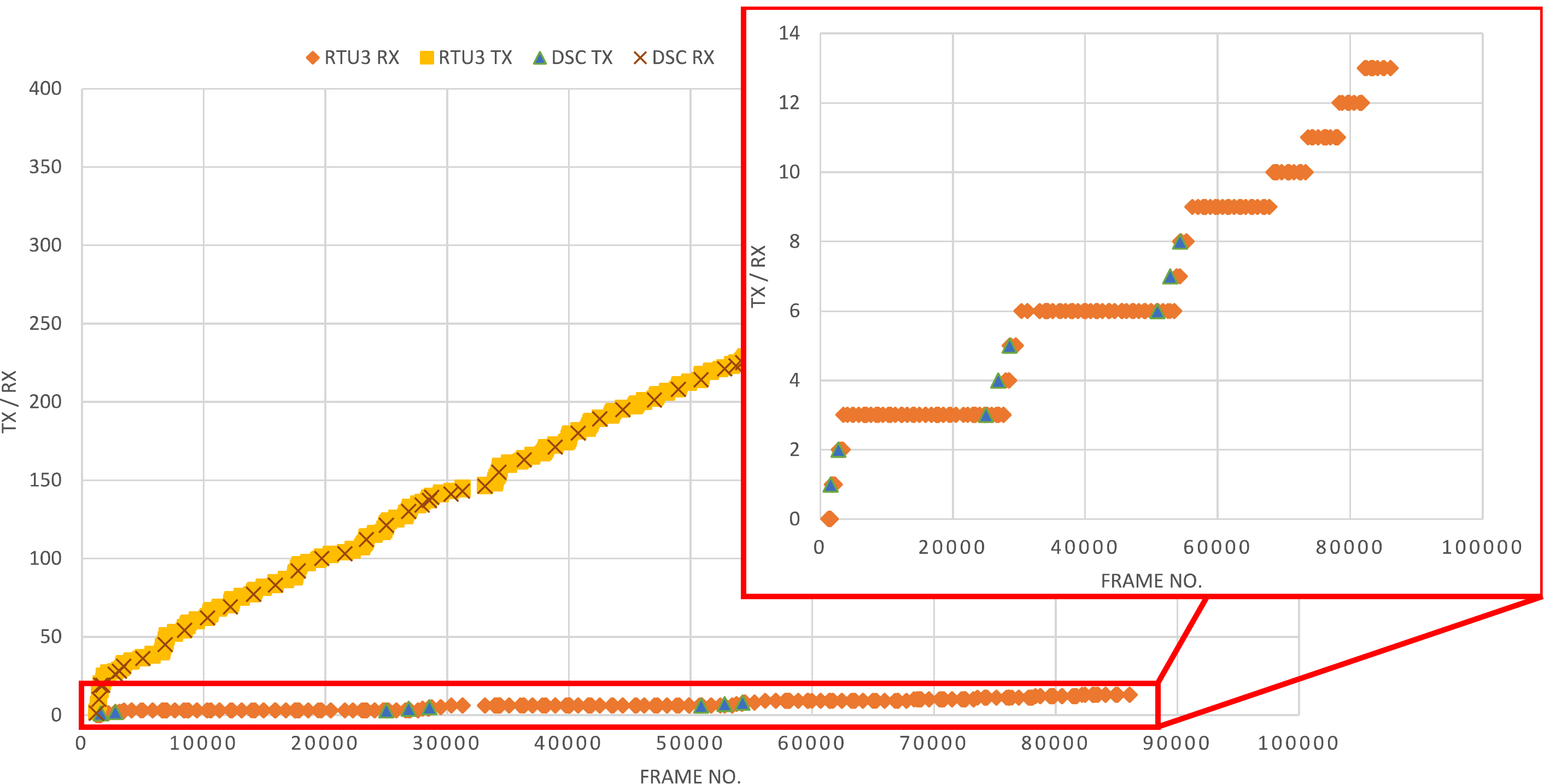}}
    \caption{Incremental sequence numbers of transmitted (TX) and received (RX) I-frames between \ac{rtu} 3 and \ac{dsc} indicating inconsistencies in tracked exchanged \ac{asdu} packets within the sequence numbers.}
    \label{fig:result_plot_txrx}
    \vspace{-1em}
\end{figure}
Using \ac{mitm}-attack tool for \ac{iec104}, we performed the experiment outlined in Section~\ref{subsec:testbed_expirement} and now present and discuss the corresponding experiment results.
The sample data recorded during the experiment, i.e., active power and voltage measurements at the \acp{mp} and the transmitted set-point commands (marked with corresponding normalized set-point at trigger times as defined in Section~\ref{subsec:testbed_expirement}), is depicted in Figure~\ref{fig:result_plot_measurements}. %
As these results show, our \ac{mitm} tool can successfully inject control commands to the respective \acp{rtu} via the \ac{mitm} agents.
Thus, the \ac{fdi} actions instructed by the \ac{mitm} coordinator affect the electrical network of our \ac{sg} laboratory in the same way as normal control commands sent by the \ac{dsc}, but with reversed operational goals (e.g., counteracting the operation of \ac{dsc}).
In the following analysis, we investigate the possibilities of detecting \ac{fdi} using a sophisticated \ac{mitm} attack and exploit these insights for the development of process-aware \ac{ids}.
In the communication traces recorded at the network tap of Switch 1, we can observe the control commands sent by the \ac{dsc}, but not those sent by the \ac{mitm} tool.
However, depending on the technical sophistication of the implementation, other indicators of the attack can potentially be extracted from the traces in addition to meta-analysis of communication behavior such as \ac{rtt} and MAC/IP/port (in)consistencies.
For example, in our conducted experiments, we observed inconsistencies in MAC addresses between received packets from \ac{rtu} 3 and sent packets from \ac{dsc} in the recorded communication traffic within the process network. %
In terms of new network participants, we can also observe new inbound and outbound traffic generated by the \ac{c2} infrastructure of the built attack scenario by observing new IP addresses in the process network (cf. Figure~\ref{fig:result_plot_c2}). %
In addition, we can also observe the effect of \ac{mitm} interception on the communication latency measured with the \ac{rtt}, with clear outliers observed with \ac{mitm} deployment (cf. Figure~\ref{fig:result_plot_rtt}).
Furthermore, anomalies in the communication flow may be observed based on the \ac{iec104} standard, which also dictates the communication behavior of, e.g., the \ac{iec104} server endpoint to return the received control commands to the \ac{iec104} client endpoint after successful processing with a \ac{cot} of \ac{actcon} or \ac{actterm}.
For example, as shown in Figure~\ref{fig:result_plot_cmdflow}, in our experiment, we can observe that \ac{actterm} packets are sent from \ac{rtu} 3 without corresponding C\_SE\_NC\_1 control commands with \ac{cot} activation sent from \ac{dsc}.
In addition, the 15-bit sequence numbers in the I- and S-formatted \ac{apdu} packets (TX, RX) can also reveal inconsistencies that can be used to indicate the \ac{mitm}-based \ac{fdi}.
The receive sequence number RX of \ac{rtu} 3 increases (approximately after 7000 frames) without the transmit number TX of \ac{dsc} increasing, i.e., no I-frames are sent, indicating suspicious behavior at the protocol application level (cf. Figure~\ref{fig:result_plot_txrx}).
Consequently, our experiments provide several indicators of \ac{mitm}-based \ac{fdi} attacks, which can be detected by, e.g., stateful monitoring of communication flow, checking the consistency of IP and MAC addresses in bilateral communications, observing noticeable latency changes in communication traffic, checking new network participants with unknown IP addresses, and sequence number inconsistencies at the protocol application layer.
In combination with plausibility and consistency checks of process data (e.g., consistency of measured values and operational plausibility of control commands), a cross-domain \ac{ids} approach with process awareness using knowledge from information and operational technology could be achieved.
The experiments are based on worst-case assumptions for \ac{fdi} via \ac{mitm}, which can be successively relaxed to present additional attacker scenarios that provide further indications building on the discussed implications of this work.
The findings presented in this paper have only emerged from a specific analysis of the communication and process data from the field experiments within our \ac{sg} testbed.

\section{Conclusion} \label{sec:conclusion}
The development and validation of appropriate countermeasures against cyber-attacks in \acp{sg} require information and data from real attack scenarios. %
To lay a foundation to provide such data, we propose an \ac{sg} laboratory environment consisting of primary and secondary technology components that enables a holistic investigation of cyber-security issues (Section~\ref{sec:testbed}).
In an effort to better understand the complex operations behind \ac{mitm}-based \ac{fdi} cyber-attacks, we have developed an \ac{mitm} tool that uses basic \ac{fdi} techniques to inject control commands into our \ac{sg} lab environment (Section~\ref{sec:attack}).
We demonstrate the applicability and provide insights into our tool and environment by performing example attack scenarios consisting of a remote coordinator controlling multiple agents (Section~\ref{sec:result}).
Our results verify the functionality of the tool we developed and also provide insights into \ac{mitm}-based \ac{fdi} attacks that can be utilized in novel \ac{ids} approaches with domain-specific and stateful detection mechanisms.
For instance, the \ac{mitm}-based \ac{mitm} was observed to produce inconsistencies in address fields within bilateral communications, application-level sequence numbers, and foreign network communications.
Future work will address the modular extension of the \ac{ict} infrastructure to achieve greater diversity in terms of the components used, such as the extension of the process network with a meshed / ring topology, segmented into edge and \ac{scada} networks.
This also includes integrating security measures such as \ac{ids}, which use the results of this work in their detection mechanism to gain further insight into the applicability and effectiveness of novel detection approaches in practical environments.

\noindent\textsc{Acknowledgments}\hspace{1em}
This work has partly been funded by the German Federal Ministry for Economic Affairs and Energy (BMWi) under project funding reference 0350028.

\end{document}